\documentclass[aps,preprint,showpacs,amsmath,amssymb]{revtex4}
\usepackage{epsfig}
\usepackage{wasysym}
\usepackage{amssymb}

\begin{document}

\newcommand{\bvec}[1]{\mbox{\boldmath ${#1}$}}
\title{Origin of the second peak in the cross section
       of the $K^+\Lambda$ photoproduction}
\author{T. Mart and M. J. Kholili}
\affiliation{Departemen Fisika, FMIPA, Universitas Indonesia, Depok 16424, 
  Indonesia}
\date{\today}
\begin{abstract}
  By using a covariant isobar model and
  the latest experimental data we have analyzed the
  role of the $P_{13}(1900)$ and $D_{13}(2080)$ resonances in the kaon
  photoproduction process $\gamma p\to K^+\Lambda$. Special attention
  has been paid to the region where the second peak in the cross section
  is located, i.e. at total c.m. energies around 
  $1.9$ GeV. It is found that this peak originates
  mostly from the $P_{13}(1900)$ resonance contribution. Although the 
  contribution of the $D_{13}(2080)$ resonance is not negligible,
  it is much smaller than that of the $P_{13}(1900)$ state. 
  Our finding
  confirms that the $P_{13}(1900)$ resonance is also important in explaining 
  the beam-recoil double polarization data $C_x$ and $C_z$, provided that 
  the mass and the width of this resonance 
  are 1871 and 131 MeV, respectively.
\end{abstract}
\pacs{13.60.Le, 25.20.Lj, 14.20.Gk}

\maketitle

In 1998 the SAPHIR Collaboration observed for the first time a 
structure in the cross section of the $\gamma p\to K^+\Lambda$
process at a total c.m. energy $W\approx 1.9$ 
GeV \cite{Tran:1998qw}. This structure was analyzed and interpreted
as evidence for a ``missing'' $D_{13}(1895)$ resonance 
\cite{missing-d13} in the model called Kaon-Maid \cite{kaon-maid}.
In spite of the fact that the inclusion of this resonance
significantly improves the agreement between model predictions
and experimental data, different interpretations have been also 
proposed \cite{Janssen:2001pe,MartinezTorres:2009cw}.  
Nevertheless, despite
considerable efforts devoted to settle this issue, there has been
no solid answer to the question: which resonance or mechanism
is responsible for this structure? 

Armed with the new generation of kaon 
photoproduction data from the CLAS 
\cite{Bradford:2005pt,Bradford:2007}, 
LEPS  \cite{Sumihama:2005er,Hicks_2007} and 
GRAAL \cite{lleres09,lleres07} Collaborations, 
especially the double polarization 
$C_x$ and $C_z$ data from CLAS \cite{Bradford:2007}, 
the Bonn-Gatchina group reported the result of their
coupled-channels partial waves analysis, 
that the structure should 
come from the contribution of the $P_{13}(1900)$ resonance 
\cite{Nikonov:2007br}. 
To our knowledge, the possibility of the $P_{13}(1900)$ resonance 
as the origin of this structure 
was first pointed out in Ref.~\cite{missing-d13}.
Nevertheless, it was ruled out because the extracted
decay width did not agree with the prediction of the constituent
quark model \cite{capstick94}. The role of this resonance 
was also briefly discussed in Refs.
\cite{Janssen:prc,Mart:2006dk,Julia-Diaz:2006is}
and finally  in Refs. \cite{Nikonov:2007br,Sarantsev:2005tg}.
Since most analyses were performed in the framework
of partial waves, it is therefore important to check 
this finding using the same tool as in Kaon-Maid, so that
a comparison with Kaon-Maid can be made under 
the same conditions. Moreover,
more precise experimental data \cite{mcCracken} have been 
just made available after the Bonn-Gatchina report \cite{Nikonov:2007br} appeared.
Thus, we believe that a more accurate analysis could
be expected. 

To this end, we consider the standard nucleon resonances
in the Particle Data Group (PDG) listing \cite{pdg2010} which have
masses between the $K^+\Lambda$ threshold (1.609 GeV) and 2.2 GeV,
the same energy range considered in the Kaon-Maid analysis  \cite{missing-d13}.
To simplify the analysis, as well as to make a fair comparison 
with Kaon-Maid, we limit the resonance spin only up to 3/2. 
Furthermore, we also include the $P_{11}(1840)$ state, which 
was found to be important in the photoproduction of 
$K^+\Lambda$, $K^+\Sigma^0$, 
and $K^0\Sigma^+$ \cite{Sarantsev:2005tg}. 

\begin{table}[b]
  \centering
  \caption{Parameters of three important resonances obtained from fitting 
    to the kaon photoproduction data in models 
    A and B
    compared to those obtained from refitting
    Kaon-Maid, i.e. models A1 and B1. 
    Numerical values printed with italic fonts 
    indicate that the corresponding parameters are fixed during
    the fitting process.}
  \label{tab:fit-result}
  \begin{ruledtabular}
  \begin{tabular}[c]{lrrrr}
    Parameter & \multicolumn{2}{c}{Present work} &
    \multicolumn{2}{c}{Kaon-Maid}\\
    \cline{2-3}\cline{4-5}
    & A & B & A1 & B1 \\
    \hline
    $m_{D_{13}(2080)}$ (MeV)      &1886       &{\it 2080} &1976 &{\it 2080}\\
    $\Gamma_{D_{13}(2080)}$ (MeV) & 244       &{\it 450}  &736  &{\it 450}\\
    $G^{(1)}_{D_{13}(2080)}$ &$-0.176$   &$0.098$    &0.809  & 0.325\\
    $G^{(2)}_{D_{13}(2080)}$ &$-0.085$   &$0.015$    &0.726  & 0.244\\
    $m_{P_{13}(1900)}$ (MeV)      &{\it 1900} &1871   &{\it 1900} & 1954\\
    $\Gamma_{P_{13}(1900)}$ (MeV) &{\it 180}  &131     &{\it 180} & 123\\
    $G^{(1)}_{P_{13}(1900)}$ &$-0.012$   & $ 0.009$ & $0.026$ &$-0.025$\\
    $G^{(2)}_{P_{13}(1900)}$ &$-0.326$   & $-0.203$ & $0.038$ &$-0.266$ \\
    $m_{P_{11}(1840)}$ (MeV)      & 1952  & 1843    &-&-\\
    $\Gamma_{P_{11}(1840)}$ (MeV) & 413   &  311    &-&-\\
    $g_{P_{11}(1840)}$            & 0.583 & 0.661   &-&-\\
    [1ex]
    \hline
    $N_{\rm data}$                &3566 &3566 & 319  & 319 \\
    $\chi^2/N_{\rm dof}$          &2.57 &2.68 & 2.42 & 3.12\\
  \end{tabular}
  \end{ruledtabular}
\end{table}

Our covariant isobar model is constructed from the appropriate
Feynman diagrams consisting of the background and resonance terms
with hadronic form factors inserted in hadronic vertices
\cite{Haberzettl:1998eq}.
The background terms 
consist of the standard $s$-, $u$-, and $t$-channel Born terms 
along with the $K^{*+}(892)$ and $K_1(1270)$ $t$-channel vector 
mesons. Two hyperon resonances that have been found to be 
important in reducing the divergence of the Born terms 
at high energies \cite{Janssen:2001pe}, the
$S_{01}(1800)$ and $P_{01}(1810)$, are also included.
For the resonance terms the model takes the $S_{11}(1650)$,  
$D_{13}(1700)$, 
$P_{11}(1710)$, $P_{13}(1720)$, $P_{11}(1840)$, $P_{13}(1900)$, 
$D_{13}(2080)$, $S_{11}(2090)$, and  $P_{11}(2100)$ nucleon resonances
into account.
Their coupling constants were determined from fitting to a
database consisting of differential cross section $d\sigma/d\Omega$ 
\cite{Bradford:2005pt,mcCracken,Sumihama:2005er,Hicks_2007}, 
recoil polarization $P$ \cite{Bradford:2005pt,lleres07},
beam-recoil double polarization $C_x,C_z$ \cite{Bradford:2007}
and $O_{x'},O_{z'}$ \cite{lleres09},
as well as photon $\Sigma$ and target $T$ asymmetries \cite{lleres07}
data.
Thus, our present database consists of 3566 data points, whereas
Kaon-Maid was only fitted to 319 data points \cite{Tran:1998qw}.

\begin{figure}[b]
  \begin{center}
    \leavevmode
    \epsfig{figure=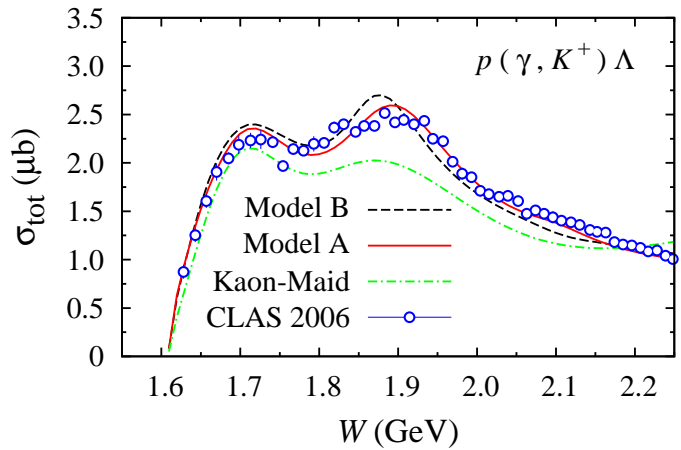,width=100mm}
    \caption{(Color online) Total cross sections obtained by fitting
    the mass and width of the $P_{13}(1900)$ (dashed line) and
    $D_{13}(2080)$ (solid line) resonances compared with that obtained
    from Kaon-Maid (dash-dotted line) \cite{kaon-maid}.
    Experimental data are from the CLAS Collaboration 
    \cite{Bradford:2005pt} and were not used in the
    fitting process.}
   \label{fig:kltot} 
  \end{center}
\end{figure}

To investigate the role of the $D_{13}(2080)$ and  $P_{13}(1900)$
resonances in the $\gamma p\to K^+\Lambda$ process, 
we perform two different fits. In the first fit
we fix the mass and width of the $P_{13}(1900)$ resonance
to their PDG values \cite{pdg2010}, i.e. 1900 and 180 MeV, respectively, 
whereas the mass and width of the $D_{13}(2080)$ state are taken as free
parameters. In the second fit, the mass and width of the $P_{13}(1900)$ 
resonance are considered as free parameters, whereas those of
the $D_{13}(2080)$ state are fixed to the PDG values, i.e., 2080 
and 450 MeV, respectively. For the sake of brevity, the first (second)
fit will be called Model A (B). In all fits
the mass and width of the $P_{11}(1840)$ resonance are taken as free
parameters.

\begin{figure}[t]
  \begin{center}
    \leavevmode
    \epsfig{figure=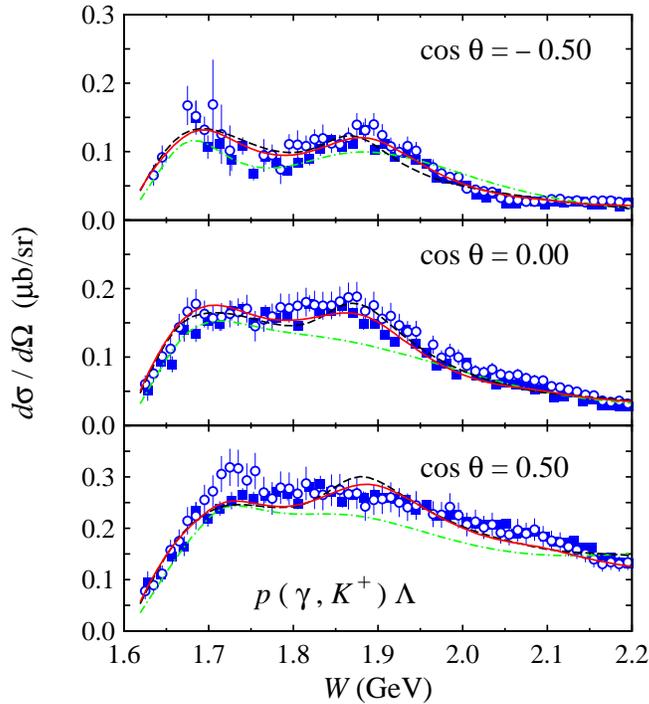,width=90mm}
    \caption{(Color online) Same as Fig.~\ref{fig:kltot}, but for the
      differential cross sections sampled at three different kaon angles. 
      Experimental data are from the 
      CLAS Collaboration (solid squares \cite{Bradford:2005pt}
      and open circles \cite{mcCracken}).}
   \label{fig:dkpl_w} 
  \end{center}
\end{figure}

\begin{figure}[b]
  \begin{center}
    \leavevmode
    \epsfig{figure=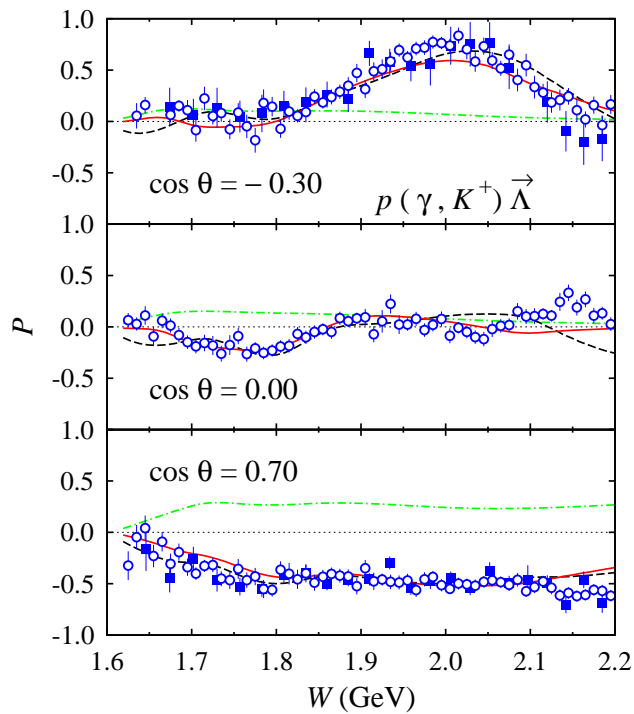,width=90mm}
    \caption{(Color online) Same as Fig.~\ref{fig:dkpl_w}, but for the
      recoil polarization observable $P$. Experimental data are from the 
      CLAS Collaboration (solid squares \cite{Bradford:2005pt}
      and open circles \cite{mcCracken}).}
   \label{fig:pollam_w} 
  \end{center}
\end{figure}

\begin{figure*}[!]
  \begin{center}
    \leavevmode
    \epsfig{figure=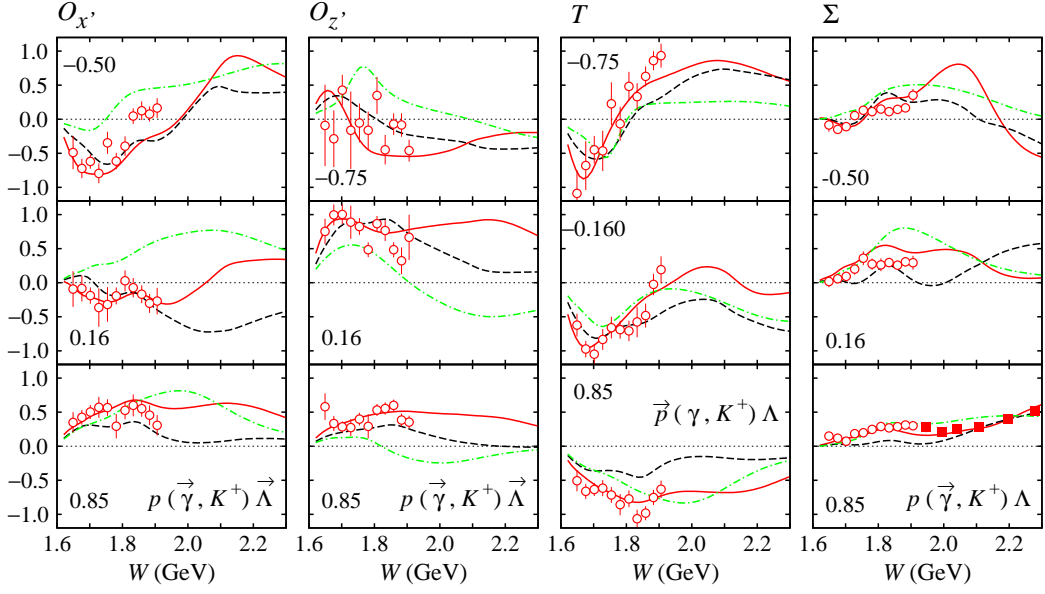,width=140mm}
    \caption{(Color online) Same as Fig.~\ref{fig:pollam_w}, but for the
      beam-recoil double polarization observable $O_{x'}$ and $O_{z'}$
      as well as the target $T$ and beam $\Sigma$ asymmetries. 
      Experimental data are from the 
      GRAAL \cite{lleres09,lleres07} (open circles) and LEPS Collaborations
      (solid circles) \cite{Sumihama:2005er}.}
   \label{fig:ox_oz_w} 
  \end{center}
\end{figure*}

\begin{figure}[!]
  \begin{center}
    \leavevmode
    \epsfig{figure=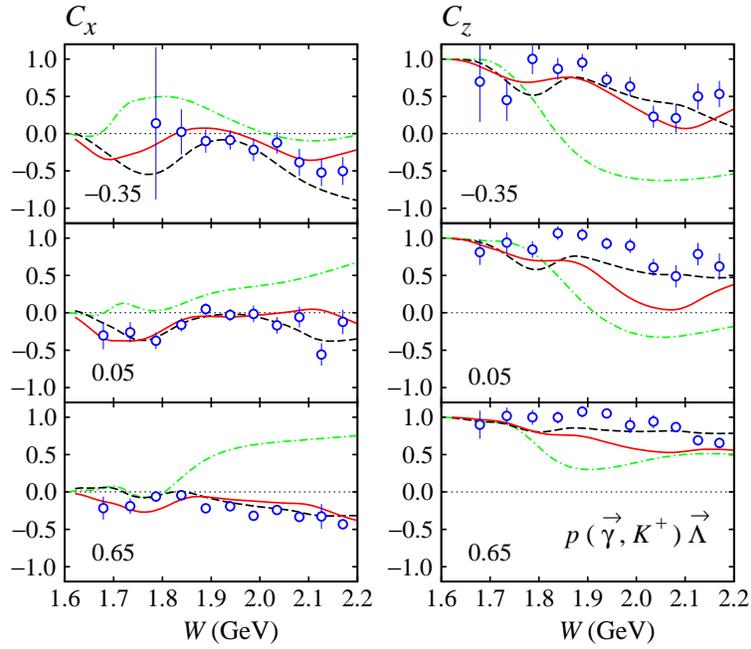,width=100mm}
    \caption{(Color online) Same as Fig.~\ref{fig:pollam_w}, but for the
      beam-recoil double polarization observable $C_x$ and $C_z$. 
      Experimental data are from the 
      CLAS Collaboration \cite{Bradford:2007}.}
   \label{fig:cx_cz_w} 
  \end{center}
\end{figure}

\begin{figure}[t]
  \begin{center}
    \leavevmode
    \epsfig{figure=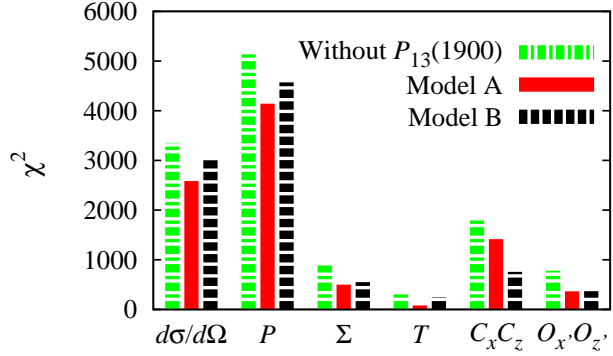,width=90mm}
    \caption{(Color online) Individual $\chi^2$ contributions from the
      differential cross section $d\sigma/d\Omega$, recoil polarization $P$,
      photon asymmetry $\Sigma$, target asymmetry $T$, and beam-recoil
      double polarization $C_x,C_z$ and $O_{x'},O_{z'}$ data.}
   \label{fig:chi2} 
  \end{center}
\end{figure}

\begin{figure}[t]
  \begin{center}
    \leavevmode
    \epsfig{figure=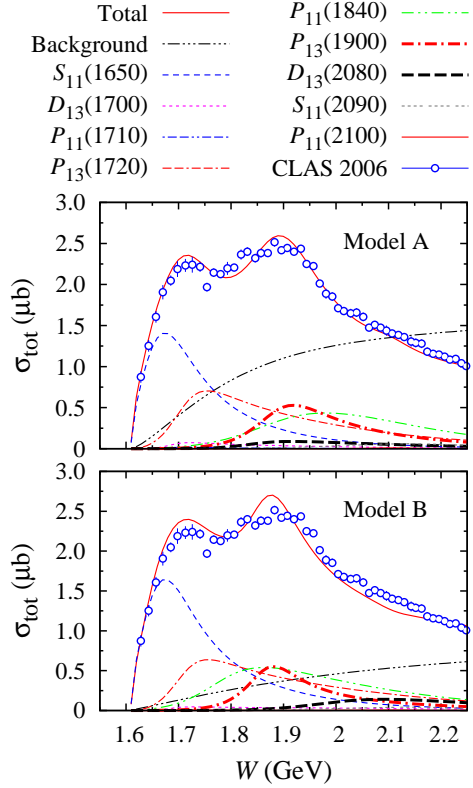,width=70mm}
    \caption{(Color online) Contribution of the background and
      resonance amplitudes to the total cross section 
      of the $\gamma p\to K^+\Lambda$ process when the mass and width 
      of the $D_{13}(2080)$ (Model A) or $P_{13}(1900)$
      (Model B) resonance are fitted. In both panels 
      contributions of the $D_{13}(2080)$ and $P_{13}(1900)$
      resonances are indicated by bold dashed and dash-dotted
      lines, respectively. 
      Experimental data are from the CLAS Collaboration 
      \cite{Bradford:2005pt}.}
   \label{fig:contrib} 
  \end{center}
\end{figure}

\begin{figure}[t]
  \begin{center}
    \leavevmode
    \epsfig{figure=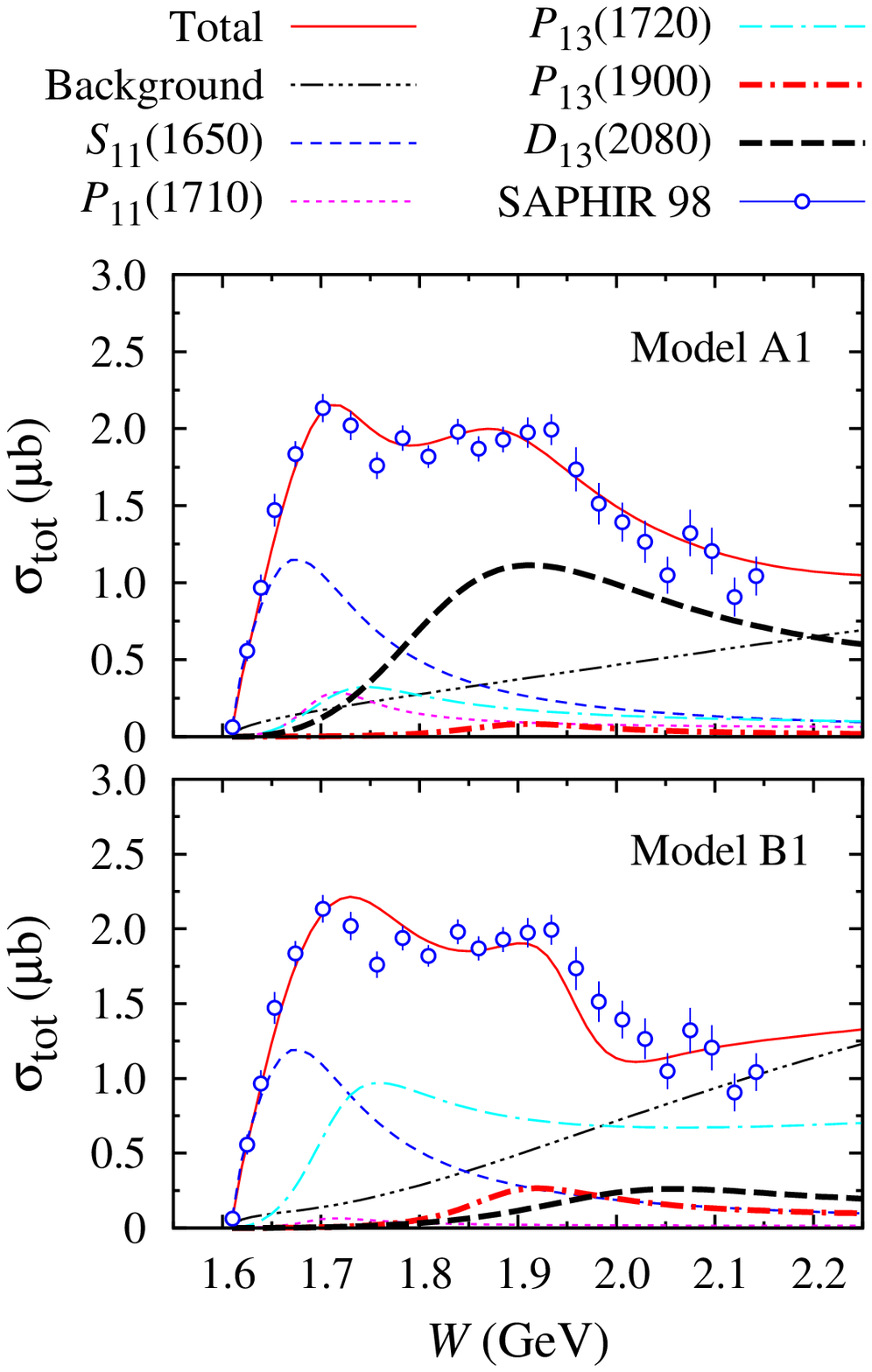,width=70mm}
    \caption{(Color online) Same as Fig.~\ref{fig:contrib}, but for the 
      refitted Kaon-Maid model. Note that the number of nucleon
      resonances used in the Kaon-Maid model is different from that 
      of the present work.
      Experimental data are from the SAPHIR 
      Collaboration \cite{Tran:1998qw}.}
   \label{fig:kmaid_contrib} 
  \end{center}
\end{figure}

Table \ref{tab:fit-result} shows the parameters of three 
most important resonances extracted from the fitting process. 
Obviously, the fitted $D_{13}$ and $P_{13}$ masses tend to 
have values around 1900 MeV. This result might
indicate that both $D_{13}$ and $P_{13}$ states could 
significantly contribute in both models. We note that when we
exclude the $P_{13}(1900)$ resonance the 
best $\chi^2/N_{\rm dof}$ obtained is 3.52, which is significantly 
larger than that obtained from both models.

It has been found that the $P_{13}(1900)$ resonance is quite important
in reproducing the $C_x$ and $C_z$ data \cite{Nikonov:2007br}.
In the present analysis we found that without this 
resonance, the contribution of the 
$\chi^2$ from the $C_x$ and $C_z$ data to the total $\chi^2$ 
is about 15\%. Including this resonance in Model A (B)
increases (decreases) this number to 16\% (8\%). 
The latter emphasizes the role of the $P_{13}(1900)$
state in explaining the $C_x$ and $C_z$ data, provided that
the mass and width of this state are taken as free parameters. 
Presumably, this is due to the structure shown by the 
$C_x$ and $C_z$ data at $W$ slightly below 1.9 GeV
(see Fig.~\ref{fig:cx_cz_w}), which can be better
explained by a $P_{13}$ resonance rather than a $D_{13}$ one.

However, it should be remembered that the 
increase of the $\chi^2$ contribution
after including the $P_{13}(1900)$ resonance in Model A does not
mean that the $P_{13}(1900)$ is insignificant in explaining the
the $C_x$ and $C_z$ data in this model, 
since the relative contribution discussed above
refers to the total $\chi^2$, which is certainly smaller in Model A 
(i.e. 2.57 as compared to 3.52).
This is elucidated by the individual $\chi^2$ contributions 
shown in Fig.~\ref{fig:chi2}.
Without the $P_{13}(1900)$ resonance the $C_x$ and $C_z$ data
contributes $\chi^2\approx 1832$ or equivalent to $\chi^2/N\approx 9$,
where $N=202$ is the number of  $C_x$ and $C_z$ data with 
total c.m. energies up to 2.2 GeV.
Including this resonance in  Model A 
(B) results in $\chi^2\approx 1414~(756)$ or 
$\chi^2/N\approx 7~(4)$. Thus, one could conclude that 
in both models the role of the $P_{13}(1900)$ state is 
found to be important in explaining the $C_x$ and $C_z$ data, 
especially in Model B. 
However,  Fig.~\ref{fig:chi2} also indicates that only 
the $C_x$ and $C_z$ data prefer Model B, in which the  $P_{13}$
mass is 1871 MeV. 
Therefore, 
the second peak in the cross section (as well as other observables
except the  $C_x$ and $C_z$ ones) prefer a "different" $P_{13}$
resonance with a mass of about 1900 MeV (Model A). This is
understandable since the position of the peak is 
located around 1900 MeV.
The result also explains the shift
of the second peak calculated using Model B from the data, as is
obviously seen in Figs.~\ref{fig:kltot} and \ref{fig:contrib}.

The need for two different $P_{13}$ resonances in order to explain
the experimental data around 1900 MeV could indicate the existence of
two $P_{13}$ resonances with masses around 1900 MeV. Indeed, 
in their recent study the Bonn-Gatchina group \cite{Nikonov:2007br} 
found two poles around 1900 MeV, as will be discussed below.

The performance of the two models in explaining experimental data
is shown in Figs.~\ref{fig:kltot}--\ref{fig:cx_cz_w}, where we also
display predictions of Kaon-Maid for comparison. The underprediction
of Kaon-Maid in both total and differential cross sections is understandable,
since the SAPHIR 1998 cross sections \cite{Tran:1998qw} are smaller than
the CLAS ones \cite{Bradford:2005pt}, especially at the second peak
around 1.9 GeV. The better agreement of Model A with experimental
data can be observed in all but $C_x$ and $C_z$ data, which is 
directly understood from the individual $\chi^2$ contributions shown in
Fig.~\ref{fig:chi2}. In fact, in both total and differential 
cross sections shown in 
Figs.~\ref{fig:kltot} and \ref{fig:dkpl_w} 
the second peak predicted by Model B
seems to be shifted from experimental data, which might lead us to 
conclude that the second peak originates from the $D_{13}$
contribution. However, this is not true.

To analyze the individual contributions of
nucleon resonances to this process, we plot contributions
of each resonance to the total cross section for both models 
in Fig.~\ref{fig:contrib}. Obviously, contributions of the $S_{11}(1650)$
and $P_{13}(1720)$ resonances explain the first peak of the cross section.
It is also clear that the $P_{11}(1710)$ resonance does not 
show up in this figure due to its small coupling to this process. 
This result 
corroborates our previous finding that uses the multipoles
formalism to describe nucleon resonances \cite{Mart:2006dk}. 
The absence of the $P_{11}(1710)$ resonance has been also pointed out in 
an extended partial-wave analysis of $\pi N$ scattering data 
\cite{Arndt:2006bf}.

Obviously, Fig.~\ref{fig:contrib} shows that 
the $P_{13}(1900)$ resonance is responsible for the
second peak in both models, whereas contribution of the $D_{13}(2080)$ 
state at this point is relatively small. 
This finding is in good agreement with the claim 
of the Bonn-Gatchina group \cite{Nikonov:2007br}, which 
found two poles located at 1870 
and 1950 MeV. Clearly, our finding corresponds to the first pole
(see the second column of Table \ref{tab:fit-result}). 
Furthermore, our result is also consistent with the previous 
coupled-channels study \cite{Julia-Diaz:2006is} and 
a very recent
kaon photo- and electroproduction study based on a single-channel
covariant isobar model \cite{maxwell}. As shown in Table II of
Ref.~\cite{maxwell}, the magnitude of the  $P_{13}(1900)$ coupling
constants is substantially larger than that of the $D_{13}(2080)$ 
ones. This is valid not only for fitting to photoproduction data,
but also for fitting to the combination of photo- and
electroproduction data. Since the $P_{13}$ and $D_{13}$ resonances
have different parities, we have checked the result of Ref.~\cite{maxwell}
explicitly and found that the contribution of the $P_{13}$ 
state is much larger
than that of the $D_{13}$ state.

It also appears from Table \ref{tab:fit-result} 
that both models yield different values of 
the $P_{11}(1840)$ mass. Model B gives a better
agreement with Ref.~\cite{Sarantsev:2005tg}, whereas the
extracted mass in Model A seems to be too high.
Nevertheless, we also note that the later analysis from
the Bonn-Gatchina group \cite{anisovich-EPJA}
yields a slightly larger mass range,
i.e. 1850-1880 MeV. 

If both $P_{13}$ and $D_{13}$ masses and widths are simultaneously 
fitted then we find a result almost similar to Model A, 
except the mass of the $P_{13}$ is slightly shifted from 1900 MeV to 1891 
MeV. Furthermore, it is also understood that the important role of the 
$P_{13}(1900)$ in explaining the $C_x$ and $C_z$ data could be 
interpreted as simulating the final state interactions that are
sensitive to the $C_x$ and $C_z$ observables. Therefore, although
the present result corroborates the finding of the coupled-channels 
work of Ref.~\cite{Nikonov:2007br}, a more thorough study using a 
dynamical coupled-channels approach, which fully takes into 
account the final state interaction effects, is still required.

The finding presented in this paper is obviously in contrast to the 
conclusion drawn more than a decade ago on the 
evidence of the $D_{13}(1895)$ resonance \cite{missing-d13}. 
Perhaps, it is interesting to ask why such a conclusion 
could be drawn. There are two possible answers to this 
question. The first one corresponds to the criteria of the "missing 
resonance." In Kaon-Maid the SAPHIR data were fitted to some possible
states with masses around 1900 MeV found in a constituent quark model
\cite{capstick94}, i.e., the $S_{11}$, $P_{11}$, $P_{13}$, and $D_{13}$
resonances.
The extracted masses of these states are found to be 
$1847$, $1934$, $1853$, and $1895$ MeV, with the 
corresponding $\chi^2/N_{\rm dof}=2.70, 3.29, 3.15$, and $3.36$, 
respectively. However, instead of using the $\chi^2$, the relevant 
"missing resonance" was determined
by matching the corresponding decay width, which can be directly calculated 
from the extracted coupling constants, with the prediction of 
the constituent
quark model \cite{capstick94}. As a result, the $D_{13}$ state 
was found to be the most relevant "missing resonance".

The second answer is related to experimental data. As discussed above, 
the use of the $P_{13}$ "missing resonance" 
to describe the SAPHIR data results in 
$\chi^2/N_{\rm dof}=3.15$, which is 
substantially larger than the use of the $S_{11}$  "missing resonance".
This indicates that to produce the second peak 
the data prefer an $S_{11}$ state rather than a $P_{13}$ state.
To further investigate the role of the $P_{13}(1900)$ resonance
in the Kaon-Maid model,  
we refit the original model, but including this
state in addition to the $D_{13}$ state in the fit, 
and using the same database as in the original model. 
The relevant extracted parameters are listed in the fourth and fifth
columns of Table~\ref{tab:fit-result}, while the contributions
of individual resonances are depicted in Fig.~\ref{fig:kmaid_contrib}.
The result indicates that if we allow the $D_{13}$ mass to vary, while the
$P_{13}$ mass is fixed to 1900 MeV (model A1), then  the $D_{13}$ contribution
 will dominate the whole cross section and simultaneously
build up the second peak. Contribution of the $P_{13}(1900)$ state is found
to be tiny. Such a result is clearly still consistent with Kaon-Maid results.

However, a different conclusion would be obtained if we kept the 
$D_{13}$ mass fixed at 2080 MeV and varied the $P_{13}$ mass
in the fit (model B1). As shown in  Fig.~\ref{fig:kmaid_contrib},
contribution of the $D_{13}$ state 
is strongly suppressed now, whereas contribution
of the $P_{13}$ state is only slightly increased. To produce the second peak, 
contributions of both resonances must be added by a larger background.
As a result, the total cross section obtained using this method shows
a substantial difference from the previous one, especially 
at $W\approx 2.0$ GeV. Therefore, we may conclude that the 
SAPHIR 1998 data do not prefer a $P_{13}$ state as 
a dominant contributor to the second peak in the cross section.

In conclusion we have analyzed the role of $P_{13}(1900)$ 
and $D_{13}(2080)$ resonances in the $K^+\Lambda$ photoproduction
off a proton, focusing on the second peak in the cross section
as well as on the CLAS $C_x$ and $C_z$ data. We found that the peak
originates mostly from the $P_{13}(1900)$ resonance. In contrast
to Kaon-Maid results, the contribution of the $D_{13}(2080)$ is much smaller,
even though its mass was fitted and found to be 1886 MeV, i.e.,
very close to the position of second peak.
The $P_{13}$ resonance is also found to be important in 
reproducing the $C_x$ and $C_z$ data. The absence of the $P_{13}(1900)$ 
contribution in Kaon-Maid is related to the SAPHIR 1998 data, since
the corresponding second peak can be best explained by means of
the $D_{13}(2080)$ resonance. The present finding does not by any  
means reject the claim that the second peak could provide evidence
for a $D_{13}$ resonance with $m\approx 1900$ MeV. It 
only shows that the evidence is weak.

This work has been supported in part by the University of Indonesia
and the Competence Grant of the Indonesian 
Ministry of Education and Culture.

\end{document}